# A Multi-User Effective Computation Offloading Mechanism for MEC System: Batched Multi-Armed Bandits Approach

Hangfan Li[1], Xiaoxiong Zhong[1,2,*], Xinghan Wang[2], Yun Ji[1], Sheng Zhang[1]

[1]Graduate School in Shenzhen, Tsinghua University, Shenzhen 518055, China

[2]Peng Cheng Laboratory, Shenzhen, 518000, P. R. China

*Corresponding author: Xiaoxiong Zhong, email: xixzhong@gmail.com

*Abstract-* With the development of 5G technology, mobile edge computing (MEC) is becoming a useful architecture, which is envisioned as a cloud computing extension version. Users within MEC system could deal with data processing at edge terminals, which can reduce time for communication or data transmission. Multi-armed bandits (MAB) algorithms are powerful tools helping users offloading tasks to their best servers in MEC. However, as the number of users and tasks growing, the frequency of selecting servers and the cost of making decision is growing rapidly under traditional MAB algorithms. Inspired by this, in this paper, we propose a Batch-based Multi-user Server Elimination (BMSE) algorithm to solve such problem, which includes two sub-algorithms. We firstly propose a sub-algorithm in user level (BMSE-UL) to reduce the time cost. In BMSE-UL, users can simplify its own available server groups and offload tasks collectively. Then another sub-algorithm in system level (BMSE-SL) is proposed to reduce the frequency of making decision. In BMSE-SL, the system can cut down all the suboptimal task offloading actions and make the choosing option unique. Furthermore, we establish the optimality of the proposed algorithms by proving the sub-linearity convergence of their regrets and demonstrate the effectiveness of BMSE by extensive experiments.

*Index Terms –*mobile edge computing (MEC), batch learning, multi-armed bandit (MAB)

## I. INTRODUCTION

With the development of the fifth generation (5G), the number of communications between individuals is growing rapidly all over the world. Cloud computing has been unable to meet the requirements of task computing or transmission under certain circumstances. Mobile edge computing (MEC) has been presented to address these issues, which can bring storage and computation resources closer to edge devices. Computation task will be offloaded for execution at edge device, which can reduce delay. Due to the limited computing resources of edge devices, the servers are unable to provide unlimited computational offload services for all tasks among edge devices. Therefore, the effective task offloading mechanism and the maximization of system performance is a challenging issue.

The strategies to address issues is designed by algorithms, like multi-armed bandits (MAB) framework. The traditional MAB algorithms including Bayesian-based Thompson Sampling (TS) and non-Bayesian-based Upper Confidence Bound (UCB1) are wildly used by not only MEC system, but also other communications framework or other fields [1], [2], [3], [4], [8], [9]. In [5], TS has been used in routing, another implementation in communications. In [17], tasks are breakdown into smaller subtasks which are easy to reduce transmission latency. In [18], MAB based FA-UCB algorithm is proposed to deal with uncertainty and asymmetry of network state information. In [19], coexisting users are becoming research targets and decentralized task offloading strategies DEBO is proposed to achieve a close-to-optimal service performance in MEC system. In [20] and [22], online learning policies based on adversary MAB framework are proposed to deal with peer and flows competition. In [21], MAB based SAGE algorithm is proposed to offer a better performance under different quality of service requirements (QoS) in MEC system. And in [10], virtual machines or servers in abundance are used to execute edge computing tasks and Multi-Agent MAB based CB-UCB and DB-UCB are proposed to minimizing task computing delay in MEC system. In [23], an MAB based algorithm is proposed to deal with the uncertainty of MEC system considering energy-efficient and delay-sensitive for a dynamic environment MEC system. All mentioned workings are based on traditional MAB which might cause some problems. The traditional MAB algorithms could only help MEC system make one task offload decision each round. Although the cost of making decision is difficult to analysis, the more options for choosing the more cost to handle. If the number of options is large, the expense on making decision might be huge. Batch learning is a good way helping system decreasing the number of decision-making. Mathematically, batch-based MAB algorithms like batch-TS and batch-UCB has good convergence when analysis regret [6], [7]. Based on current existing working, research on applying batch based MAB algorithm on MEC system is deficient.

Inspired by these, in this paper, we propose a Batch-based Multi-user Server Elimination (BMSE) algorithm to reduce the

frequency of making decision and the total time cost. The contributions of this article are as follows:
1) For multi-user computation offloading in MEC system, where the number of users is much larger than the number of servers (at least 3 times), we proposed a BMSE algorithm framework, which is based on batch MAB. BMSE contains two sub-algorithms: 1) BMSE-UL, which is used to make progress in user level and aim to reduce the number of options, and BMSE-SL, which is used to make progress in system level and aim to reduce the time cost of the whole system.
2) We rigorously establish the optimality of both sub-algorithms by using Hoeffding's inequality to reach the sub-linearity of their regrets, which is converged under upper bounded by $O(\sqrt{I^2 T \log T})$ and $O(\frac{4\sigma_{max}^2 \ln T(1-\frac{2}{T})}{\sum_{j \in A''} \Delta_j})$ respectively, achieving sub-linear convergence compared to the time horizon $T$, where $I$ is the number of users, $\sigma_{max}$ is the max length of the transmission capacity interval among users, $\Delta_j$ is average cost difference between the optimal action and action $a_j$ ( $j$ is action ID) and $A''$ is the action pool.
3) The extensive experiments demonstrate the effectiveness of BMSE, in which has around 33% of time saving with almost the same optimal rate as Multi-user UCB1 algorithm under 10000 time slots and the number of decisions making by BMSE is negligible even at 1000 time slots.

The remainder of this paper is organized as follows. In Section II, we propose the MEC system model and build the problems to be optimized. Then in Section III, we propose BMSE algorithm and do regret convergence analysis in mathematical. In the end, we evaluate the performance of the proposed scheme and provide illustrative results in Section IV and conclude the paper in Section V.

## II. SYSTEM MODEL AND PROBLEM FORMULATION

In this section, we will firstly introduce the multi-servers computing system based on MEC system. Then we will describe delay model, and present BMSE framework.

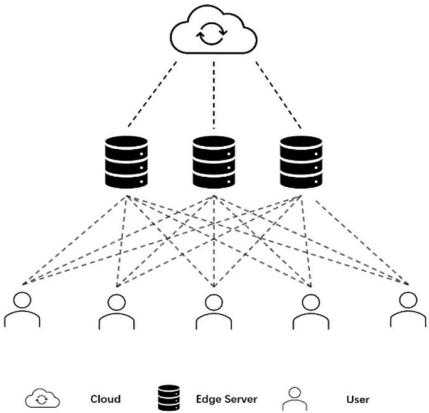

Fig. 1. System model

### 1. System model

MEC is a latency sensitive and computation intensive network. Assume that edge capacity and channel gain between different users and servers are unknown which might be the most essential element influencing transmission cost. We consider an online learning based three-layer task-offloading mobile edge computing network with a set of finite user $i$ collected in $U = \{u_1, ..., u_i, ..., u_I\}$, a set of finite Edge server $e$ collected in $EServer = \{s_1, ..., s_e, ..., s_E\}$, and a Cloud server $CServer = \{s_c\}$. In general, we assume that the number of devices is much bigger (many times or more) than the number of servers ($I \gg E$). In the MEC system, we consider a finite time slot (round, in other words, in the rest of this paper) $t \in \mathcal{T}\{1,2,3,...,T\}$ and each user $i$ (an edge device like smartphone) generates a computation task $\omega_{i,t}$ (generated by user $i$ at time $t$) and assign it to an execution server by some pre-deployed algorithms where $\omega_{i,t} = \{c_{i,t}^F, c_{i,t}^B, \pi_{i,t}\}$, where $c_{i,t}^F$ represents the task size, $c_{i,t}^B$ represents the result size and $\pi_{i,t}$ represents an execution method (including local device and all available edge servers) which will be decided by the pre-deployed algorithms.

### 2. Task generation & processing

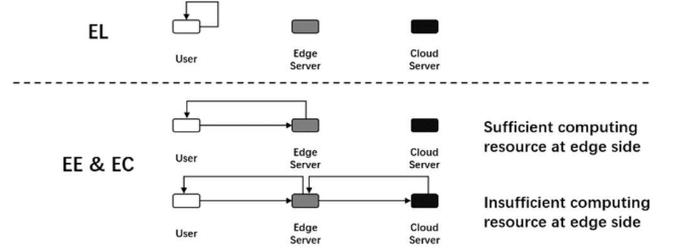

Fig. 2. Task execution patterns.

As shown in Fig .2, we define three task execution patterns: Executing Locally (EL), Executing at Edge (EE) and Executing at Cloud (EC). In pattern EL, the tasks will be executed by device CPU. In pattern EE, the task will be firstly uploaded to an edge server and user gets a result from edge server. And in pattern EC, the task will be firstly uploaded to the edge server and due to the limited computing resource, then uploaded to the cloud server and get a result. To be simplicity, $\pi_{i,t} \in \{EL\} \cup \{EServer\} \cup \{ECloud\}$, where $\{EL\}$ contains one element, $\{EServer\}$ contains $|E|$ elements and $\{ECloud\}$ contains one element. Based on the three mentioned patterns, the execution cost (delay) of task $\omega_{i,t}$ is denoted as follow:

$$D(\omega_{i,t}) = \begin{cases} D^{EL}(\omega_{i,t}), & \pi_{i,t} \in \{EL\} \\ D^{EE}(\omega_{i,t}), & \pi_{i,t} \in \{EServer\} \\ D^{EC}(\omega_{i,t}), & \pi_{i,t} \in \{ECloud\} \end{cases} \quad (1)$$

In pattern EL, the cost $D^{EL}(\omega_{i,t})$ is defined:

$$D^{EL}(\omega_{i,t}) = \frac{c_{i,t}^F}{v_i} \quad (2)$$

where $v_i$ represents the CPU capacity and could be defined as follow:

$$v_i = \frac{f_i}{\beta} \quad (3)$$

where $f_i$ represents the CPU clock frequency of user $i$ and $\beta$ represents average CPU cycles needed for executing one bit. In pattern EE and EL, the execution method $\pi_{i,t}$ is a positive integer, which represents one of available edge server, $\pi_{i,t} \in \{EE\}$ and the cost $D^{EE}(\omega_{i,t})$ contains three steps: upload to edge server $D_{i \to \pi_{i,t}}$, execute the task $D_{\pi_{i,t}}$ and get the result $D_{\pi_{i,t} \to i}$.

$$D^{EE}(\omega_{i,t}) = D_{i \to \pi_{i,t}} + D_{\pi_{i,t}} + D_{\pi_{i,t} \to i} \quad (4)$$

$$D_{i \to \pi_{i,t}} = \frac{c_{i,t}^F}{r_{i,\pi_{i,t}}} \quad (5)$$

$$D_{\pi_{i,t}} = \frac{c_{i,t}^F}{v_{\pi_{i,t}}} \quad (6)$$

$$D_{\pi_{i,t} \to i} = \frac{c_{i,t}^B}{r_{i,\pi_{i,t}}} \quad (7)$$

$$r_{i,\pi_{i,t}} = W \log_2\left(1 + \frac{p h_{i,\pi_{i,t}}}{N}\right) \quad (8)$$

The task execution capacity of server $\pi_{i,t}$ is denoted as $v_{\pi_{i,t}}$, transmission rate between user $i$ and selected edge server $\pi_{i,t}$ is denoted as $r_{i,\pi_{i,t}}$, channel bandwidth is denoted as $W$, power consumption for transmitting the input data of task is denoted as $p$, channel gain between the user and edge server is denoted as $h_{i,\pi_{i,t}}$ and the white noise power level is denoted as $N$. The task handle cost $D^{EC}(\omega_{i,t})$ containing five steps: upload task from user $i$ to edge server $\pi_{i,t}$ $D_{i \to \pi_{i,t}}$, upload task from edge server $\pi_{i,t}$ to cloud server $s_c$ $D_{\pi_{i,t} \to c}$, execute the task at cloud server $D_c$, get the result from cloud server to edge server $D_{c \to \pi_{i,t}}$ and get the result from edge server to user $D_{\pi_{i,t} \to i}$.

$$D^{EC}(\omega_{i,t}) = D_{i \to \pi_{i,t}} + D_{\pi_{i,t} \to c} + D_c + D_{c \to \pi_{i,t}} + D_{\pi_{i,t} \to i} \quad (9)$$

$$D_{i \to \pi_{i,t}} + D_{\pi_{i,t} \to c} = \frac{c_{i,t}^F}{r_{i,\pi_{i,t}}} + \frac{c_{i,t}^F}{r_{\pi_{i,t},c}} \quad (10)$$

$$D_c = \frac{c_{i,t}^F}{v_c} \quad (11)$$

$$D_{c \to \pi_{i,t}} + D_{\pi_{i,t} \to i} = \frac{c_{i,t}^B}{r_{\pi_{i,t},c}} + \frac{c_{i,t}^B}{r_{i,\pi_{i,t}}} \quad (12)$$

where $r_{\pi_{i,t},c}$ represent the transmission rate between selected edge server and cloud server which is defined by Shannon equation as the same way of $r_{i,\pi_{i,t}}$ and $v_c$ represent the task execution capacity of cloud server.

### 3. Problem formulation

The execution cost for a single user can be written as:

$$\sum_{t=1}^T D(\omega_{i,t}) \quad (13)$$

Our target is to minimize the task execution cost in MEC system. The problem is formally described in mathematical as follows:

$$\boldsymbol{P1}: minimize \sum_{i=1}^I \sum_{t=1}^T D(\omega_{i,t}) \quad (14)$$
$$s.t. \pi_{i,t} \in [EL, EE, EC]$$
$$t \in [1, T]$$
$$i \in [1, I]$$

The **P1** is non-trivial and not detailed. This is because the channel conditions and computing resource change over time and we cannot predict or find its changing rules. In addition, **P1** is also computationally prohibitive due to the large amount of optimization variables.

To improve the accessibility of the **P1**, we propose to reformulate the problem by adopting MAB framework and apply MAB algorithms which is suitable to solve task offloading strategy problems in such unpredictable MEC system. It is effective to learn the channel condition distribution within an Exploration (select an ill-considered edge server with low choosing times) and Exploitation (select an empirically well performed edge server) trade off. Assume $\pi_t = \sum_{i=1}^I \pi_{i,t}$ represent an action at time slot $t$ (action and action pool will be explained in Chapter Algorithm) and $\pi = \sum_{t=1}^T \pi_t$ represent the task offload strategy for the whole MEC system. Assume $D^*$ represent the least cost under the optimal action and denote total regret of MEC system to:

$$R_\pi = \mathbb{E}\left[\sum_{t=1}^T (D_{\pi_t} - D^*)\right] = \mathbb{E}\left[\sum_{t=1}^T D_{\pi_t} - TD^*\right] \quad (15)$$

Table 1. Notation and corresponding description.

| Variable | Description |
|---|---|
| $U = \{u_1, ..., u_i, ..., u_I\}$ | a set of users. total number is I |
| $EServer = \{s_1, ..., s_e, ..., s_E\}$ | a set of edge server. total number is E |
| $CServer = \{s_c\}$ | a cloud server. |
| $t \in \mathcal{T}\{1, 2, 3, ..., T\}$ | time slot (round) horizon. total number is T |
| $\omega_{i,t} = \{c_{i,t}^F, c_{i,t}^B, \pi_{i,t}\}$ | task generated by user $i$ at time slot $t$. |
| $c_{i,t}^F, c_{i,t}^B$ | upload task size and result task size of task $\omega_{i,t}$ |
| $\pi_{i,t}$ | execution method of task $\omega_{i,t}$. $\pi_{i,t}$ can select EL, EE, EC |
| $D(\omega_{i,t})$ | delay (cost) of task $\omega_{i,t}$ |
| $D^{EL}(\omega_{i,t}), D^{EE}(\omega_{i,t}), D^{EC}(\omega_{i,t})$ | delay of task when $\pi_{i,t}$ select EL, EE, EC respectively |
| $D_{a \to b}$ | transmission delay from a to b |
| $r_{a,b}$ | transmission rate between a and b |
| $W$ | channel bandwidth |
| $v$ | task execution capacity |
| $D^*$ | the least cost under the optimal action |
| $R_\pi$ | total regret of MEC system under strategy $\pi$ |
| $A = \{\alpha_1, ..., \alpha_J\}$ | an action pool. Total action number is J |
| $(\alpha_{j,1\sim2})_{J \times 2}$ | corresponding matric of action pool |
| $B$ | batch number |
| $\xi$ | correction parameters, a positive number |
| $\Phi_1, ..., \Phi_i, ..., \Phi_I$ | active edge server groups of each user |
| $(\bar{a}_{i,e})_{I \times E}, (\bar{b}_{i,e})_{I \times E}$ | two matric recording average cost and selecting times |
| $\sigma_{max}$ | The max length of the transmission capacity interval among users |

Obviously, due to the deficiency of essential information, we reformulate the problem as:

$$P2: \min_{\pi} R_\pi \quad (16)$$

$\pi_t$ has high dependency on offloading strategy (or policy, algorithm) and our aim is to find a suboptimal strategy achieving sub-linear regret:

$$\lim_{T \to \infty} \frac{R_\pi}{T} = 0 \quad (17)$$

## III. BMSE ALGORITHM FRAMEWORK

To solve **P2**, we will propose well performed algorithms under centralized setting. At first, we consider a traditional stochastic multi-armed bandit problem and deploy UCB1 algorithm in MEC system. As shown in the Fig.3 (Traditional MAB), for every single time slot, the system selecting an action which is a set of $(user, method)$ tuples including offloading strategy for all users. The decision cost (selecting an action in action pool by deployed algorithm) is negligible in normal MEC. However, it might be a massive expense if the size of action pool is large. Therefore, algorithms combining the batch learning and traditional MAB algorithm, are called batch MAB. The difference between Traditional MAB and Batch MAB is shown in Fig. 3.

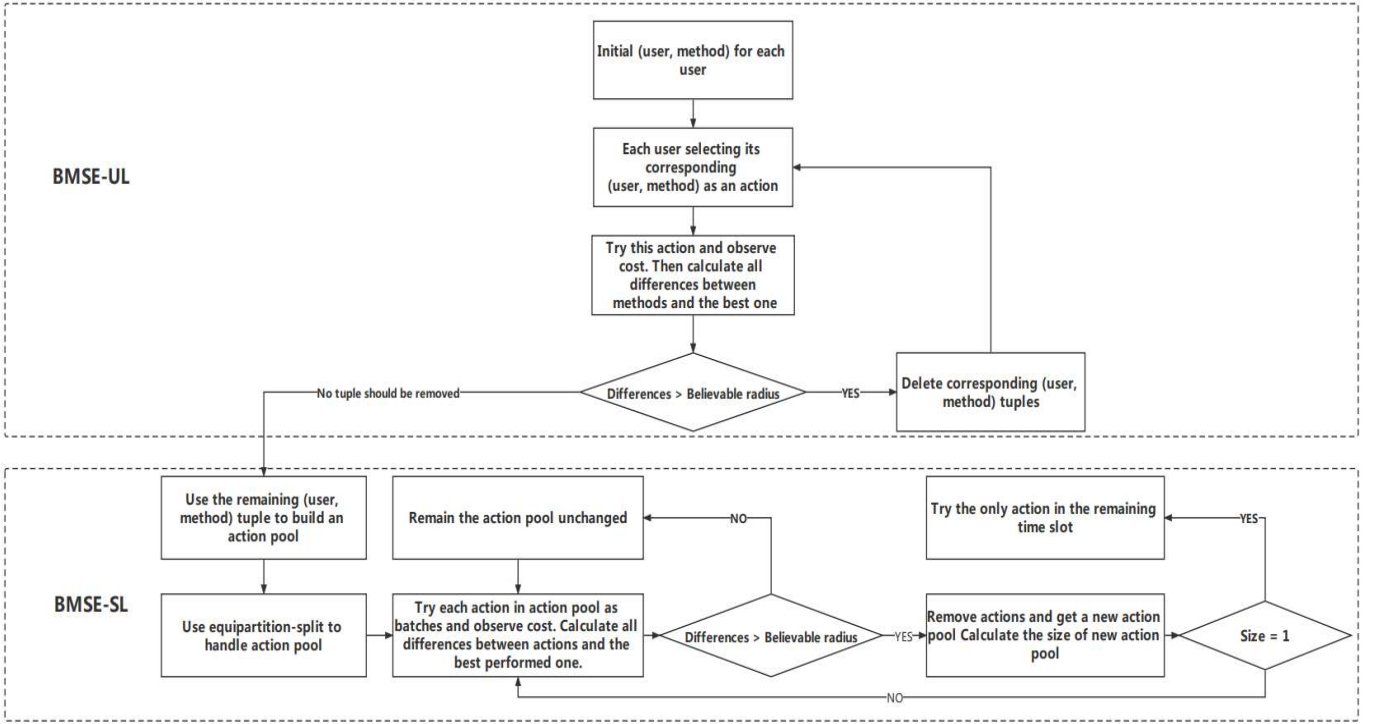

Fig 4. BMSE algorithm framework

Based on batch MAB, we propose a Batch-based Multi-user Server Elimination (BMSE) algorithm, including BMSE contains BMSE-User Level (BMSE-UL) algorithm and BMSE-System Level (BMSE-SL) algorithm, as shown in Fig. 4. There are two targets for the BMSE algorithm: 1) Hold a high-level accuracy rate by taking much less decision than Multi-user UCB1. 2) Have a lower delay by reducing the size of action pool. BMSE-UL is called making progress in BMSE user level. The main target of this algorithm is to eliminate the method for each user with obvious awful performance. Then the time will be saved by avoid wasting computing recourses on actions containing these methods. BMSE-SL is called making progress in BMSE system level. The main target for BMSE-SL is to eliminate the suboptimal actions until finding the optimal one. After there are only one action in action pool, the system does not need to make any decision. Next, we will first explain the two algorithms in details and analyze the convergence when applying them independently.

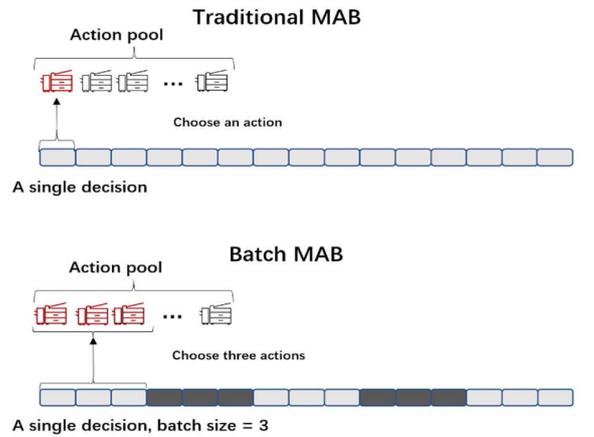

Fig. 3. Difference between Traditional MAB & Batch MAB.

Before introducing the algorithms, we give some notations. Due to the number of method of each user is defined by local execution and offloading to edge server, we denote suboptimal methods of each user are denoted to $\Phi_1, \ldots, \Phi_i, \ldots, \Phi_I$ and the initial size is $|E + 1|$ and build two $|I| \times |E + 1|$-dimensional matrix $(\bar{a}_{i,e})_{I \times (E+1)}$ and $(b_{i,e})_{I \times (E+1)}$ where $\bar{a}_{i,e}$ represents the current average cost of user $i$ executing task by method $e$ and $b_{i,e}$ represents the current selecting times of method $e$ by user $i$. Denote an action is a set of $(user, method)$ tuples, an action pool $A = \{\alpha_1, \ldots, \alpha_J\}$ contains all possible task offloading actions where the size of action pool is upper bound to $O(|E + 1|^{|I|})$ and a corresponding $J \times 2$-dimensional matrix $(\alpha_{j,1}, \alpha_{j,2})_{J \times 2}$ where $\alpha_{j,1}$ represents the current average cost of action $\alpha_j$ and $\alpha_{j,2}$ represent the current selecting times of action $\alpha_j$. $a^*$ represents the best action with least cost.

### 1. Multi-user UCB1 algorithm

For easy understanding, firstly, we introduce Multi-user UCB1 algorithm in MEC system, as shown in **Algorithm 1**. In lines 1~4 is the initial exploration stage. At the beginning $|A'|$ times, the algorithm tries each action $a_j$ in action pool and observe its cost. In lines 5~13 is the exploration and exploitation stage. The following $T - |A'|$ times is the main part for UCB1 algorithm where in lines 6~9 is the decision-making stage. The main purpose of this stage is to work out next task offloading strategy where $\xi$ is a stable correction parameter. Lines 10~13 is aimed to offload the strategy and update necessary arguments. According to the pseudo-code of Multi-user UCB1, the complexity of **Algorithm 1** is $O(n^2)$. The performance of Multi-user UCB1 is bad especially under a large size of action pool. Therefore, we propose BMSE-UL algorithm to reduce the size of action pool.

| Algorithm 1 Multi-User UCB1 |
|---|
| **Input:** |
|   The MEC system |
|   Setting Action pool $A' \leftarrow A$ |
| **Output:** |
|   The optimal action |
| 1. **for** $t = 1: \|A'\|$ **do:** |
| 2.   Try each action in $A'$ and observe action delay |
| $$d(\alpha_j) = \sum_{i=1}^{J} D(\omega_{i,t})$$ |
| 3.   Update $\alpha_{j,1} = d(\alpha_j)$ and $\alpha_{j,2} = 1$ |
| 4. **end for** |
| 5. **for** $t = (\|A'\| + 1): T$ **do** |
| 6.   **for** each action $\alpha_j$ in $\|A'\|$ **do** |
| 7.     Estimate the cost of action $\hat{\alpha}_{j,1} = \alpha_{j,1} + \sqrt{\frac{\xi \ln t}{\alpha_{j,2}}}$ |
| 8.   **end for** |
| 9.   Choose the current best performed action |
| $$\alpha_j^* = \arg \min_{\alpha_j \in A'} \hat{\alpha}_{j,1}$$ |
| 10. Observe action delay $d(\alpha_j^*)$ |
| 10. Update $\alpha_{j,1}^* = \alpha_{j,1}^* + \frac{\alpha_{j,1}^* + d(\alpha_j^*)}{\alpha_{j,2}^*}$ |
| 11. Update $\alpha_{j,2}^* = \alpha_{j,2}^* + 1$ |
| 13. **end for** |

### 2. BMSE-UL algorithm

In BMSE-UL algorithm, as shown in **Algorithm 2**, the aim is to get a set of reasonable method groups by inputting all necessary information. In lines 1, we initialize the system by dislocation choice. For instance, if method number is 3 and user number is 4. The start method index is (1, 2, 3, 1), then (2, 3, 1, 2) and (3, 1, 2, 3) is over. In lines 2 and 19~21, we run a while loop and set a shutdown condition. In lines 3~6, we run the best performance method of each user according to the pervious information and combing as an action. Try this action and obvious the cost. Lines 7~18 is the most important. We first identify the one-method users and set them choice the last method forever where the other methods are eliminated. By doing this, we can reduce the $|E + 1|$ in $(|E + 1|)^{|I|}$. Then we identify method with obviously bad performance. If the difference between methods and the best one is smaller than a believable radius $\sqrt{\frac{\xi \ln b_{i,e}}{b_i^*}}$, we call this method reasonable, otherwise, eliminate it for reasonable method group. By doing this, we can reduce $|I|$ in $(|E + 1|)^{|I|}$.

However, although we make progress on reducing size of action pool through **Algorithm 2**, we still need to make plenty of decisions. Therefore, we propose BMSE-SL algorithm to reduce the times of making decisions.

| Algorithm 2 BMSE-UL |
|---|
| **Input:** |
|   The MEC system |
|   User $U' \leftarrow User$ and user-serrver matrices $(a_{i,e})_{I \times E}$ and $(b_{i,e})_{I \times E}$ |
|   Initial method group of each user $\Phi_i' \leftarrow \Phi_i$ |
| 1. Initialize: each user $i$ dislocation choice methods in $\Phi_i'$ and combining as actions until cover all method for all users, try these actions and update the average cost |
| 2. **while** True **do** |
| 3.   Each user choice the most well-performance method in $\Phi_i'$ and combining as an action |
| 4.   Try this action and observe delay $a_{i,e}$ for each user |
| 5.   Update $b_{i,e} = b_{i,e} + 1$ |
| 6.   Update $\bar{a}_{i,e} = a_{i,e} + \frac{\bar{a}_{i,e} \times (b_{i,e}-1) + a_{i,e}}{b_{i,e}}$ |
| 7.   **for** $user$ in $U'$ **do** |
| 8.     **if** $\|\Phi_i'\| = 1$ **then** |
| 9.       Remove current $user$ from $U'$ and set the method in $\Phi_i'$ as the best |
| 10.      Pass |
| 11.     **else** |
| 12.       **for** $method$ in $\Phi_i'$ **do** |
| 13.         **if** $a_{i,e} > a_i^* + \sqrt{\frac{\xi \ln b_{i,e}}{b_i^*}}$ **then** |
| 14.           Remove current $method$ from $\Phi_i'$ |
| 15.         **end if** |
| 16.       **end for** |
| 17.     **end if** |
| 18.   **end for** |
| 19.   **if** no $method$ removed **then** |
| 20.     Break |
| 21.   **end if** |
| 22. **end while** |

## 3. BMSE-SL algorithm

In batch mode, the player commits a sequence of actions and observes the rewards after all actions in that sequence are pulled like [13]. In MEC system, we divide the time horizon $T$ into $B$ batches. It should be noted that if B = T, there is nothing different from the traditional MAB which is learning sensitive and well researched. If B = 1, then no learning can happen. The regime 1<B<T is our research interests. In BMSE-SL, the task offloading policy within a batch is determined before the task generating and this process corresponding a sequence of actions. Batch size in BMSE-SL is fixed before running the algorithm which can be designed as geometric grids like [14] where the beginning batch size can be $\Theta\left(T^{\frac{1}{B}}\right)$ or $\Theta\left(T^{\frac{1}{2-2^{1-B}}}\right)$. And it is necessary to have $B = O\left(\frac{\log T}{\log \log T}\right)$ to achieve optimal regret [11] and $B = O(\log T)$ in [12]. Based on above, we set the batch size equal to the action pool size.

In BMSE-SL, we firstly use equipartition-split to handle reasonable method groups and build an action pool. Then we eliminated suboptimal actions based on batching learning. In line 1, we propose a equipartition-split method to split the users into $|E|$ parts. For instance, split 9 users into 3 group corresponding at most $C_9^3 \times C_6^3 \times C_3^3 = 1960$ actions which is much smaller than $3^9 = 19683$. We priority split uses with less method in its method group. By doing this, the total number of actions might smaller. In [15], the placement of the edge server needs to consider the trade-off between energy consumption and task transmission delay and point out that areas with high user density should place more edge servers, vice versa. The process of user within same user density area emergence that requires task offloading conforms to Poisson distribution. Based on above truth, it has high probability action pool $A''$ contain the optimal action.

**Theorem 1.** The size of action pool by equipartition-split is upper bound by $\Theta\left(\frac{|I|!}{\left(\left\lfloor\frac{|I|}{|E|}\right\rfloor!\right)^{|E|}}\right)$, where $I$ is the number of users and $E$ is the number of servers.

*Proof:* we have the following cases:
**Case 1:** $I$ can be divided by $E$:
The number of users in each group $\mathcal{M}$ is denoted as:

$$\mathcal{M} = \frac{I}{E} \tag{18}$$

The size of action pool built by equipartition-split is:

$$C_I^{\mathcal{M}} \times C_{I-\mathcal{M}}^{\mathcal{M}} \times \cdots \times C_{I-(E-1)\mathcal{M}}^{\mathcal{M}} = \prod_{e=1}^{E} C_{I-(e-1)\times\mathcal{M}}^{\mathcal{M}} = \frac{I!}{(\mathcal{M}!)^E} \tag{19}$$

**Case 2:** $I$ cannot be divided by $E$:
Group size is $\left\lfloor\frac{I}{E}\right\rfloor$ and its number is denoted as $u$. Group size is $\left\lceil\frac{I}{E}\right\rceil$ its number is denoted as $v$:

$$u = E \times \left\lceil\frac{I}{E}\right\rceil - I \tag{20}$$

$$v = I - E \times \left\lfloor\frac{I}{E}\right\rfloor \tag{21}$$

The size of action pool built by equipartition-split is:

$$C_I^{\left\lfloor\frac{I}{E}\right\rfloor} \times \cdots \times C_{I-(u-1)\times\left\lfloor\frac{I}{E}\right\rfloor}^{\left\lfloor\frac{I}{E}\right\rfloor} \times C_{I-u\times\left\lfloor\frac{I}{E}\right\rfloor}^{\left\lceil\frac{I}{E}\right\rceil} \cdots \times C_{I-u\times\left\lfloor\frac{I}{E}\right\rfloor-(v-1)\left\lceil\frac{I}{E}\right\rceil}^{\left\lceil\frac{I}{E}\right\rceil}$$

$$= \prod_{u=1}^{u} C_{I-(u-1)\times\left\lfloor\frac{I}{E}\right\rfloor}^{\left\lfloor\frac{I}{E}\right\rfloor} \times \prod_{v=1}^{v} C_{I-u\times\left\lfloor\frac{I}{E}\right\rfloor-(v-1)\left\lceil\frac{I}{E}\right\rceil}^{\left\lceil\frac{I}{E}\right\rceil}$$

$$= \frac{I!}{\left(\left\lfloor\frac{I}{E}\right\rfloor!\right)^u \times \left(\left\lceil\frac{I}{E}\right\rceil!\right)^v}$$

$$\leq \frac{I!}{\left(\left\lfloor\frac{I}{E}\right\rfloor!\right)^u \times \left(\left\lfloor\frac{I}{E}\right\rfloor!\right)^v}$$

$$= \frac{I!}{\left(\left\lfloor\frac{I}{E}\right\rfloor!\right)^{u-v}} = \frac{I!}{\left(\left\lfloor\frac{I}{E}\right\rfloor!\right)^{E\left(\left\lceil\frac{I}{E}\right\rceil - \left\lfloor\frac{I}{E}\right\rfloor\right)}} = \frac{I!}{\left(\left\lfloor\frac{I}{E}\right\rfloor!\right)^E} \tag{22}$$

which completes the proof.

By using this method, we can initialize the batch size to $|A''|$ and batch number $B = \left[\frac{T}{|A''|}\right]$. In each batch, users offload up to $q_b = \left\lfloor\frac{B}{|A''|}\right\rfloor$ tasks where $|A''|$ will decrease due to the action elimination. In lines 2~5 represent the **initial exploration stage**. The algorithm tries each action and observe its cost in the first batch. In lines 6~23 is the **exploration and exploitation stage.** In lines 7~10, we set a shutdown condition. If only one action exists in action pool, we set it as the optimal action. The following batches are the main part for BMSE-SL where in lines 15~21 is the **decision-making stage.** At the beginning of main loop, the algorithm tries each remaining actions in action pool and update corresponding average cost. Then in decision-making stage, the algorithm eliminates worse actions and calculate next average task offloading number $q_b$. In lines 24~26, we just try the only action in action pool and no need to make decision. According to the pseudo-code in **Algorithm 2**, it is easy to obtain the complexity of **Algorithm 2**, $O(n^2)$. By applying this algorithm, the number of making decision is dropping rapidly.

---
**Algorithm 3 BMSE-SL**
---
**Input:**
 The MEC system
 Reasonable method group of each user $\Phi'_i$
1. Simplify Action pool $A'' \leftarrow A$ by equipartition-split
2. **for** $b = 1$ **do**
3.   Try each action in $A''$ and observe action delay

$$d(a_j) = \sum_{i=1}^{J} D(\omega_{i,t})$$

4.   Initialize $a_{j,1} = d(a_j)$ and $a_{j,2} = 1$
5. **end for**
6. **for** $b = 2: B - 1$ **do**
7.   **if** $|A''| = 1$ **then**
8.     Set the last action as the optimal action
8.     Break
9.   **end if**
10.  Try each action in $A''$ for $q_b$ times and observe action delay

$$d(a_j) = \sum_{i=1}^{J} D(\omega_{i,t})$$

        drop the remaining time slots
12.   **for** $a_j$ in $|A''|$ **do**
13.     Update $a_{j,1} = a_{j,1} + \frac{a_{j,1}-d(a_j)}{a_{j,2}+1}$ for $q_b$ times
14.     Update $a_{j,2} = a_{j,2} + 1$
15.   **end for**
16.   $a_j^* = \arg\min_{a_j \in A'} \hat{a}_{j,1}$
17.   **for** $a_j$ in $|A''|$ **do**
18.     **if** $a_{j,1} > a_{j,1}^* + \sqrt{\frac{\xi \ln b |A''|}{a_{j,2}^*}}$ **then**
19.       Remove $a_j$ from $A''$
20.     **end if**
21.   **end for**
22.   Update $q_b = \left\lfloor \frac{B}{|A''|} \right\rfloor$
23. **end for**
22. **for** the remaining time slot **do**
25.   Choose the optimal action in $|A''|$
26. **end for**

BMSE-UL could help us saving time by reducing the size of action pool and BMSE-SL could help us reducing the number of making decision. In BMSE algorithm, system will firstly obey BMSE-UL algorithm outputting a reasonable, small action pool and then obey BMSE-SL algorithm to reducing number of making decision by batch learning. And it is easy to work out that the complexity of BMSE is also $O(n^2)$.

**4. Regret analysis**

**Theorem 2.** The expected regret of algorithm BMSE-UL for traditional stochastic multi-armed bandits is upper bounded by $O(\sqrt{I^2 T \log T})$, where $T$ is the time horizon, and $I$ is the number of users.

*Proof:* Recall Hoeffding's inequality [16], Let $X_1, X_2, \ldots, X_n$ be independent random variables with range $[a_i, b_i]$. Let $\overline{X} = \frac{1}{n}(X_1 + \cdots + X_n)$. Then for all $\varepsilon \geq 0$:

$$\mathbb{P}\{\overline{X} - \mathbb{E}[\overline{X}] \geq \varepsilon\} \leq \exp\left(-\frac{2\varepsilon^2 n^2}{\sum_{i=1}^{n}(b_i-a_i)^2}\right) \quad (23)$$

$$\mathbb{P}\{|\overline{X} - \mathbb{E}[\overline{X}]| \geq \varepsilon\} \leq 2\exp\left(-\frac{2\varepsilon^2 n^2}{\sum_{i=1}^{n}(b_i-a_i)^2}\right) \quad (24)$$

The whole algorithm can be seen as deploying successive elimination algorithm to all users. We denote $\Delta(i, e)$ as the cost gap among arm $e$ and the largest cost gap:

$$\Delta(i)^* = \arg\max_{e \in \Phi_i} \Delta(i, e) \quad (25)$$

Assume that a single user gets its optimal state at time slot $b_{i,e}$, we obtain the regret:

$$Regret(i) = \sum_{\Phi_i} b_{i,e} \times \Delta(i, e) \quad (26)$$

According to the Hoeffding's inequality, we have

$$\Delta(i, e) \leq O(confidence\ radius) = O\left(\sqrt{\frac{\ln T}{b_{i,e}}}\right) \quad (27)$$

where $\sqrt{x}$ is a concave function and according to the Jensen's Inequality we have:

$$Regret(i) \leq \sum_{\Phi_i} b_{i,e} O\left(\sqrt{\frac{\ln}{b}}\right)$$

$$\leq \sum_{\Phi_i} O\left(\sqrt{b_{i,e} \ln T}\right)$$

$$\leq O\left(\sqrt{b_{i,e}|E| \ln T}\right)$$

$$\leq O\left(\sqrt{T \ln T}\right) \quad (28)$$

The total expected regret:

$$\sum_I Regret(i) \leq O\left(\sqrt{I^2 T \ln T}\right) \quad (29)$$

which completes the proof.

**Theorem 3.** The expected regret of algorithm BMSE-SL for batched-based stochastic multi-armed bandits is bounded by $O(\frac{4\sigma_{max}^2 \ln T\left(1-\frac{2}{T}\right)}{\sum_{j \in A''} \Delta_j})$, where $I$ is the number of users, $T$ is the time horizon $T$, $\sigma_{max}$ is the max length of the transmission capacity interval among users, $\Delta_j$ is average cost difference between the optimal action and action $a_j$ ($j$ is action ID) and $A''$ is the action pool.

*Proof:* The cost (delay) for each action in action pool $A''$ are independent random variables and its range is obtained after observing. Denote active action: for all actions exist in $A''$ at the end of each batch.

Denote correct estimation for a single action $a_j$: at the end of batch $b$, its estimated mean within $\varepsilon = \frac{\sigma_{max}}{\sqrt{2}} \times \sqrt{\frac{\ln b |A''|}{a_{j,2}}}$ of its actual mean.

$$\mathbb{P}\{correct\ estimation\} = \mathbb{P}\{\overline{X} - \mathbb{E}[\overline{X}] \leq \varepsilon\} \quad (30)$$

Denote

$$\sigma_{max} = \arg\max_{i \in I}(a_i - b_i) \quad (31)$$

Therefore, according to the Hoeffding's inequality, the probability of correct estimation any active arm at the end of any batch except the last one is:

$$\mathbb{P}\{correct\ estimation\} \geq 1 - 2\exp\left(-\frac{2\varepsilon^2 n^2}{\sum_{i=1}^{n}(b_i-a_i)^2}\right)$$

$$\geq 1 - 2\exp\left(-\frac{2\varepsilon^2 n^2}{n\sigma_{max}^2}\right)$$

$$\geq 1 - 2\exp\left(\frac{2\ln(b|A''|)\Delta_{max}^2 n^2}{2n\sigma_{max}^2 a_{j,2}}\right)$$

$$\geq 1 - 2\exp(\ln(b|A''|))$$

$$= 1 - \frac{2}{b|A''|} \quad (32)$$

The estimation of any active arm at the end of any batch before the last one is incorrect with probability at most $\frac{2}{b|A''|}$.

Hence, the total expected regret before the last batch is:

$$\mathbb{E}[Regret] = \mathbb{E}[Regret|correct\ estimations] + \mathbb{E}[Regret|\overline{correct\ estimations}] \quad (33)$$

$$\mathbb{E}[Regret] = 2 + \mathbb{E}[Regret|correct\ estimations] \quad (34)$$

Since some suboptimal actions is not eliminated before $b+1$ batch with the same selecting times as the optimal action.

Denote the average cost difference $\Delta_j$, and we have

$$\Delta_j \leq \left| \left( a_{j,1} - \frac{\sigma_{max}}{\sqrt{2}} \times \sqrt{\frac{\ln b |A''|}{a_{j,2}}} \right) \right.$$
$$\left. - \left( a_{j,1}^* + \frac{\sigma_{max}}{\sqrt{2}} \times \sqrt{\frac{\ln b |A''|}{a_{j,2}^*}} \right) \right|, j \in A''$$
$$= \left| \left( a_{j,1} - \frac{\sigma_{max}}{\sqrt{2}} \times \sqrt{\frac{\ln b |A''|}{a_{j,2}}} \right) \right.$$
$$\left. - \left( a_{j,1}^* + \frac{\sigma_{max}}{\sqrt{2}} \times \sqrt{\frac{\ln b |A''|}{a_{j,2}}} \right) \right|$$
$$\leq \frac{2\sigma_{max}}{\sqrt{2}} \sqrt{\frac{\ln b |A''|}{a_{j,2}}}$$
$$(35)$$

which is the upper bound of actions, and $a_{j,2} \leq \frac{4\sigma_{max}^2 \ln(b|A''|)}{\Delta_j^2}$.

Hence, the total expected regret can be expressed:

$$\mathbb{E}[Regret|correct\ estimations]$$
$$= \sum_{j \in A''} \Delta_j \mathbb{E}[a_{j,2}|correct\ estimations]$$
$$\leq \sum_{j \in A''} \frac{4\sigma_{max}^2 \ln(B|A''|)\left(1 - \frac{2}{B|A''|}\right)}{\Delta_j}$$
$$\leq \sum_{j \in A''} \frac{4\sigma_{max}^2 \ln T \left(1 - \frac{2}{T}\right)}{\Delta_j}$$
$$(36)$$

which completes the proof.

## IV. PERFORMANCE EVALUATION

In this section, we make six experiments. Fig. 5, Fig. 6, Fig. 7 aim to show the powerful performance of BMSE algorithm comparing to Multi-User UCB and Epsilon-Greedy algorithm. in optimal rate and total time cost dimensionality. Fig. 8 and Fig. 9 aim to show the universal applicability of BMSE algorithm. According to [19], [24], the parameter settings are shown in Table 2. The aim of Fig. 10 is to show difference of the size growing speed of action pool under BMSE algorithm and traditional MAB algorithms like Multi-User UCB when the number of users grows.

***Optimal rate*** is the ratio of optimal choosing actions to overall actions.

***Size of action pool*** represents how many actions in action pool.

Table 2. Simulation parameter settings

| Variables | Default setting |
|---|---|
| Local device CPU frequency | 3.0GHz |
| CPU cycle needed to handle one bit task | 3000 |
| Channel bandwidth | 30kHz |
| Noise power | 50W |
| Channel gain | 0.125~1.0 (0.125 correspond edge to cloud) |
| Signal power | 3200W |
| Edge capacity | 50MB/s ~ 51MB/s |
| Cloud capacity | 100GB/s |
| $\xi$ | 1 |

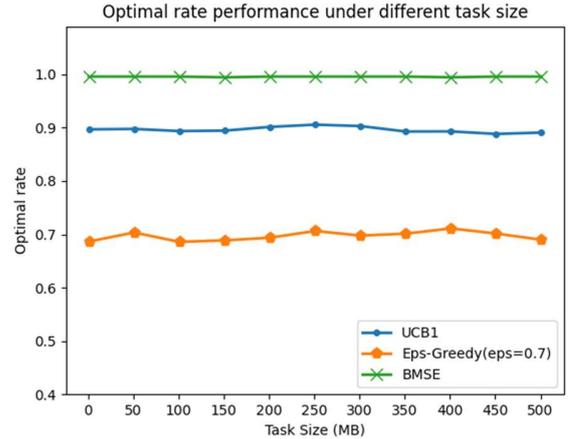

Fig. 5. Optimal rate vs Task size.

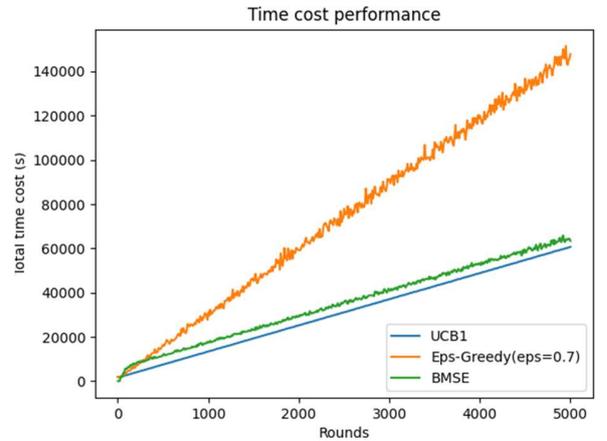

Fig. 6. Time cost vs Rounds.

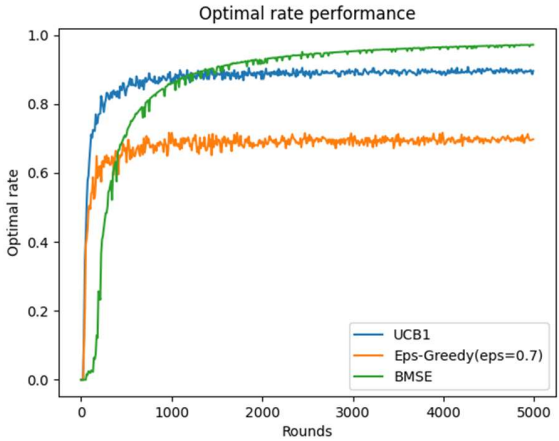

Fig. 7. Optimal rate vs Rounds.

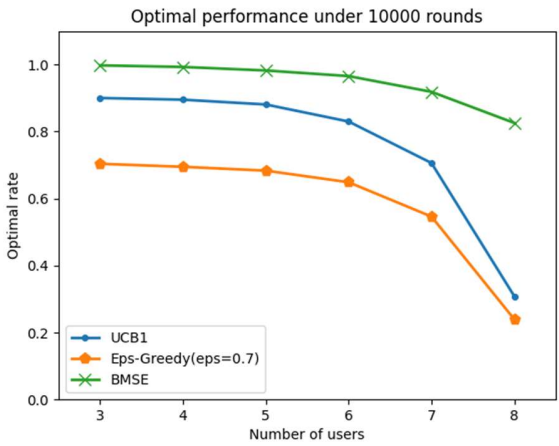

Fig. 8. Optimal rate vs Number of users.

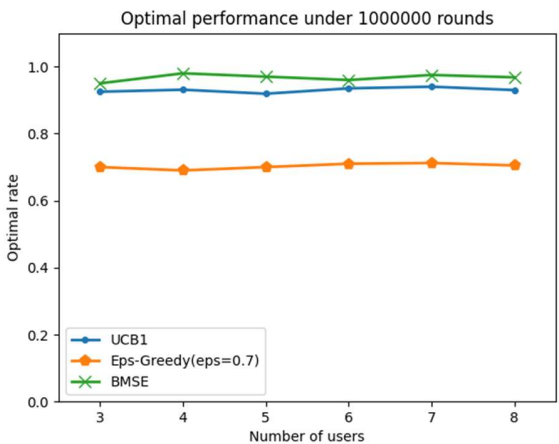

Fig. 9. Optimal rate vs Number of users.

In the Fig. 5 ~ 7, known the size of action pool is huge, we run the experiments under a MEC system with 4 users and 2 edge servers. In Fig. 5, we set rounds to five thousand to see whether the task size has influence to the algorithms. It is evident that task size could no longer enhance or weaken algorithms. In other words, BMSE algorithm is suitable for all kinds of task processing MEC system. We set task size 200 MB in Fig. 6 and Fig. 7. These two experiments are the directly manifest of powerful performance of BMSE algorithm. In Fig. 6, at the beginning stage, BMSE algorithm might spend some time on exploration, but this phase does not last long and will stay slow growth as Multi-User UCB does. In Fig. 8, we can draw a conclusion that although BMSE does not converge as fast as Multi-User UCB and Eps-Greedy, BMSE does have the highest optimal rate. Thanks to the successive elimination theory, the deficient performance actions will be discarded finally. By doing this, the performance of BMSE in optimal rate is good.

We set task size 500MB and the number of edge server two in Fig. 8 and Fig. 9 and set rounds $10^4$ and $10^6$ respectively. As the number of users growth, the size of action pool increases rapidly. Under such condition, due to the user elimination in BMSE-UL and equipartition-split in BMSE-SL, the size of action pool could be control in a reasonable range. In Fig. 9, if the number of rounds is huge or infinite, all algorithms could have a superior performance as the number of users growing. However, in Fig. 8, if the number of rounds is small or finite, BMSE could keep a superior performance as the number of users growing. Therefore, BMSE algorithm could be applied into a wider range of MEC systems.

Fig. 10 is an album of pictures which are aimed to point out the size difference between traditional MAB algorithms and BMSE when the number of users become three times to five times than the number of edge servers. It is evident that the size of action pool in BMSE algorithm is much smaller than in Multi-user UCB1. As the number of edge servers and the number of users grows, the size gap becomes more obvious. The action pool will be processed in both user level and system level where the number of available methods of each user and the number of suboptimal actions is much smaller than Multi-user UCB1. If the traditional MAB algorithm is applied, in the face of such a huge action pool, the cost might be exceptionally large.

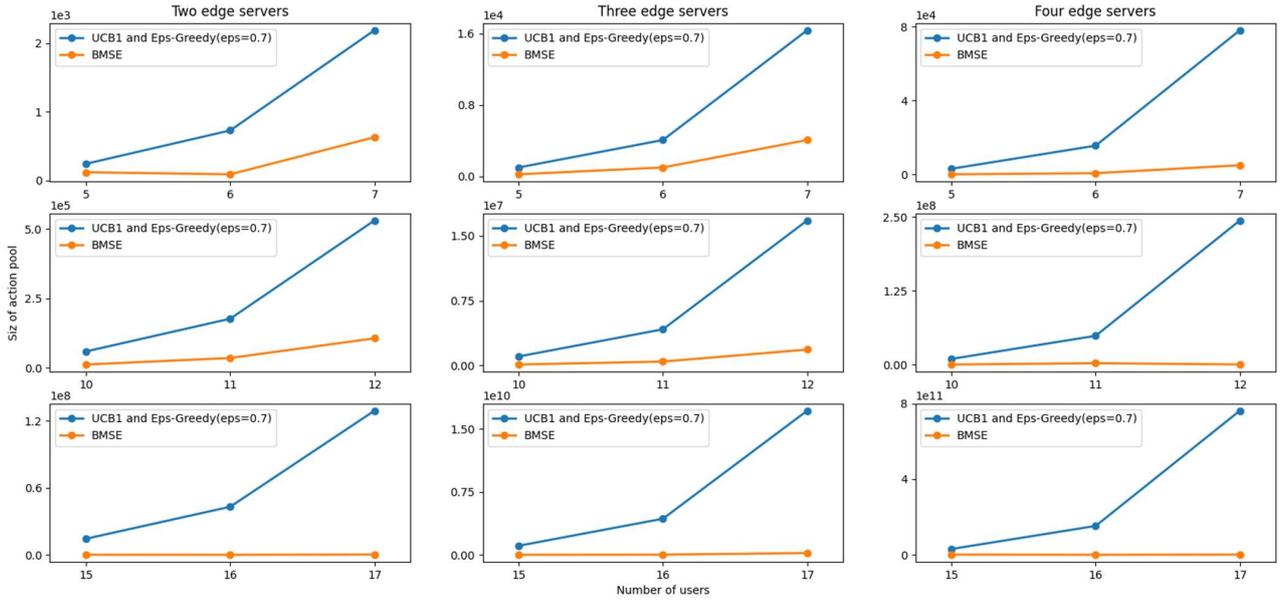

Fig. 10. The number of users vs Size of action pool (Two edge servers).

## V. CONCLUSION

In this paper, we propose a batched MAB algorithm based multi-user computation task offloading mechanism, BMSE, for the large-scale MEC system, whose goal is maximizing the resource usage, saving cost (delay) of the whole system and reducing the number of decisions. In proposed scheme, we split BMSE into two sub-algorithms, BMSE-UL and BMSE-SL. Then we analysis the regret of two sub-algorithms is converged under upper bounded by $O\left(\sqrt{I^2 T \log T}\right)$ and $O\left(\frac{4\sigma_{max}^2 \ln T \left(1-\frac{2}{T}\right)}{\sum_{j \in A''} \Delta_j}\right)$ respectively, which can achieve sub-linear convergence compared to the time horizon $T$. In the end, we conduct extensive experiments to demonstrate the effectiveness of BMSE. BMSE has higher optimal rate than Multi-user UCB1 with the same time cost as Multi-user UCB1 algorithm under 5000 rounds and BMSE could adapt to a more demanding environment than traditional MAB algorithms.

## ACKNOWLEDGE

This work was supported by the National Natural Science Foundation of China (Grant Nos. 61802221), and the Peng Cheng Laboratory Project (Grant No. PCL2021A02).

## REFERENCES

[1] Y. Sun, J. Song, S. Zhou, et.al., "Task replication for vehicular edge computing: a combinatorial multi-armed bandit based approach," in proc. of IEEE GLOBECOM 2018:1-7.
[2] Y.Sun, G. Kaan, S. Suleyman, "An online minimax optimal algorithm for adversarial multiarmed bandit problem," IEEE Trans. Neural Networks Learn. Syst. 29(11): 5565-5580, 2018.
[3] L. Chen, J. Xu, "Budget-constrained edge service provisioning with demand estimation via bandit learning," IEEE J. Sel. Areas Commun. 37(10): 2364-2376, 2019.
[4] W. Ding, T. Qin, X. Zhang, et.al., "Multi-armed bandit with budget constraint and variable costs," AAAI, 2013.
[5] Z. Huang, Y. Xu, J. Pan, "TSOR: thompson sampling-based opportunistic routing," IEEE Trans. Wirel. Commun. 20(11): 7272-7285, 2021.
[6] C. Kalkanli, A. Ozgür, "Batched thompson sampling," CoRR abs/2110.00202, 2021.
[7] A. Kolnogorov, S. Garbar, "Multi-armed bandit problem and batch ucb rule," MATH.ST, 2019.
[8] N. Merlis, S. Mannor, "Batch-size independent regret bounds for the combinatorial multi-armed bandit problem," COLT 2465-2489, 2019.
[9] Y. Mao, M. Chen, J. Pan, et.al., "A batched multi-armed bandit approach to news headline testing," CoRR abs/1908.06256, 2019.
[10] B. Wu, T. Chen, W. Ni, et.al., "Multi-agent multi-armed bandit learning for online management of edge-assisted computing," IEEE Trans. Commun. 69(12): 8188-8199, 2021.
[11] Z. Gao, Y. Han, Z. Ren, et.al., "Batched multi-armed bandits problem," NeurIPS: 501-511, 2019.
[12] H. Esfandiari, A. Karbasi, A. Mehrabian, et.al., "Batched multi-armed bandits with optimal regret," CoRR abs/1910.04959, 2019.
[13] H. Esfandiari, A. Karbasi, A. Mehrabian, et.al., "Regret bounds for batched bandits," AAAI: 7340-7348, 2021.
[14] V. Perchet, P. Rigollet, S. Chassang, et.al., "Batched bandit problems," COLT: 1456, 2015
[15] Y. Li, A. Zhou, X. Ma, et.al., "Profit-aware edge server placement," IEEE Internet Things J. 9(1): 55-67, 2022.
[16] W. Hoeffding, "Probability inequalities for sums of bounded random variables," Journal of The American Statistical Association 58 (301): 13-30, 1963.
[17] S. Misra, S. Rachuri, P. Deb, et.al., "Multiarmed-bandit-based decentralized computation offloading in fog-enabled IoT," IEEE Internet Things J. 8(12): 10010-10017, 2021.
[18] B. Wu, T. Chen, K. Yang, et.al., "Edge-centric bandit learning for task-offloading allocations in multi-RAT heterogeneous networks," IEEE Trans. Veh. Technol. 70(4): 3702-3714, 2021.
[19] X. Wang, J. Ye, J. Lui, "Decentralized task offloading in edge computing: a multi-user multi-armed bandit approach," CoRR abs/2112.11818, 2021.
[20] M. Yang, H. Zhu, H. Wang, et.al., "Peer to peer offloading with delayed feedback: an adversary bandit approach," ICASSP: 5035-5039, 2020.

[21] S. Gao, T. Yang, H. Ni, et.al., "Multi-armed bandits scheme for tasks offloading in MEC-enabled maritime communication networks," ICCC: 232-237, 2020.

[22] C. Hua, L. Wang, P. Gu, "Online offloading in dense wireless networks: an adversary multi-armed bandit approach," WCSP: 1-6, 2018.

[23] S. Choorchian, S. Maghsudi, "Multi-armed bandit for energy-efficient and delay-sensitive edge computing in dynamic networks with uncertainty," IEEE Trans. Cogn. Commun. Netw. 7(1): 279-293, 2021.

[24] N. Anh-Nhat, D. Ha, V. Vo, et.al., "Performance analysis and optimization for iot mobile edge computing networks with rf energy harvesting and uav relaying," IEEE Access 10: 21526-21540, 2022.